%% file: main.tex
\documentclass[compsoc,conference,a4paper,10pt,times]{IEEEtran}
\IEEEoverridecommandlockouts
\usepackage{cite}
\usepackage{amsmath,amssymb,amsfonts}
\usepackage{algorithmic}
\usepackage{graphicx}
\usepackage{subcaption} 
\usepackage{textcomp}
\usepackage{bmpsize}
\usepackage{lipsum}
\usepackage[colorlinks=true,urlcolor=black]{hyperref}
\def\BibTeX{{\rm B\kern-.05em{\sc i\kern-.025em b}\kern-.08em
    T\kern-.1667em\lower.7ex\hbox{E}\kern-.125emX}}

\usepackage{xspace}
\usepackage[table,xcdraw]{xcolor}
\usepackage{colortbl} 
\usepackage{multirow}

\usepackage{booktabs}
\usepackage{adjustbox}
\usepackage{tikz}

\usepackage[normalem]{ulem}  
\usepackage{xcolor}          

\newcommand{\MakeMarkups}[3]{
  \expandafter\newcommand\csname #2\endcsname[1]{\textcolor{#3}{\textbf{[##1]}}}
}

\MakeMarkups{Sina}{SA}{blue}
\MakeMarkups{Sandra}{Siby}{purple}
\MakeMarkups{Hamed}{HH}{magenta}
\MakeMarkups{Marios}{MK}{cyan}
\MakeMarkups{Mohamad}{MM}{orange} 
\definecolor{lightgreen}{rgb}{0.6, 1, 0.6}

\newcommand{\para}[1]{\smallskip \noindent \textbf{#1}}
\newcommand*\circled[1]{\tikz[baseline=(char.base)]{
            \node[shape=circle,draw,inner sep=1pt] (char) {#1};}}
\providecommand{\etal}{\emph{et al.}\xspace}            
\begin{document}

\input{abbs}

\title{An Early Experience with Confidential Computing Architecture for On-Device
Model Protection\thanks{Accepted to the 8th Workshop on System Software for Trusted Execution (SysTEX 2025).}}
\author{
  \IEEEauthorblockN{
    Sina Abdollahi\IEEEauthorrefmark{1}, 
     Mohammad Maheri\IEEEauthorrefmark{1}, 
    Sandra Siby\IEEEauthorrefmark{2},
    Marios Kogias\IEEEauthorrefmark{1}, 
     Hamed Haddadi\IEEEauthorrefmark{1}
  }
  \IEEEauthorblockA{
    \IEEEauthorrefmark{1}Imperial College London \quad
    \IEEEauthorrefmark{2}New York University Abu Dhabi \quad
  }
}
\maketitle
\input{parts/abstract}

\section{Introduction}
\input{parts/introduction}





\section{Background}
\label{background}
\input{parts/background}


\section{Framework Architecture}\label{architecture}
\input{parts/model}
\input{parts/system}

\section{Evaluation}\label{Evaluation}
\input{parts/evaluation}

\section{Discussion}
\input{parts/discussion}
\section{Related Works } \label{related-works}
\input{parts/relatedworks}
\section{Conclusion}
\input{parts/conclusion}

\section*{Acknowledgments}
\input{parts/Acknowledge}
\bibliographystyle{IEEEtran}
\bibliography{bibliography}

\appendices
\section{Experimental Setup} \label{section:Experimental Setup}
\input{parts/FVP_setup}
\section{Inference Overhead}\label{section:Inference Overhead appendix}

\input{parts/Inference_Overhead}
\section{Realm Setup Overhead}\label{section:Realm Setup Overhead appendix}
\input{parts/realm_setup_overhead}

\end{document}

%% file: abbs.tex
\newcommand\eg{\emph{e.g.},\xspace}
\newcommand\ie{\emph{i.e.},\xspace}
\newcommand\etc{\emph{etc}.\xspace}
\newcommand\via{\emph{via}}
\providecommand{\etal}{\emph{et al.}\xspace}

\newcommand{\paranew}[1]{\smallskip \noindent \textbf{#1}}
\newcommand{\parait}[1]{\smallskip \noindent \textit{#1}}

%% file: parts/abstract.tex
\begin{abstract}
Deploying machine learning (ML) models on user devices can improve privacy (by keeping data local) and reduce inference latency. Trusted Execution Environments (TEEs) are a practical solution for protecting proprietary models, yet existing TEE solutions have architectural constraints that hinder on-device model deployment. Arm Confidential Computing Architecture (CCA), a new Arm extension, addresses several of these limitations and shows promise as a secure platform for on-device ML. In this paper, we evaluate the performance–privacy trade-offs of deploying models within CCA, highlighting its potential to enable confidential and efficient ML applications.
Our evaluations show that CCA can achieve an overhead of, at most, 22\% in running models of different sizes and applications, including image classification, voice recognition, and chat assistants.  This performance overhead comes with privacy benefits, for example, our framework can successfully protect the model against membership inference attack by 8.3\% reduction in the adversary's success rate. To support further research and early adoption, we make our code and methodology publicly available.
\end{abstract}

%% file: parts/introduction.tex
 Machine-learning (ML) models are increasingly being deployed on edge devices  for various purposes such as health monitoring, anomaly detection, face recognition, voice assistants \etc Running models locally can provide low-latency services to users  without the need for sending data to an external entity. As end-users typically lack the resources required to train a model, they prefer to utilize a pre-trained, reliable model owned by a third party for inference and, potentially, personalization. The model owner, having invested significant resources in training the model, requires robust security assurances to safeguard the model’s integrity and usage. Without these guarantees, the owner may not be willing to deploy the model on end devices.

Various solutions have been proposed for model protection on the edge. Cryptographic techniques such as homomorphic encryption (HE)\cite{orlandi2007oblivious, gilad2016cryptonets, van2019sealion} or secure multiparty communication (SMC)\cite{mohassel2017secureml, riazi2018chameleon} are hindered by computational and communication overheads, while the use of trusted execution environments (TEEs) is considered a more efficient approach.  A TEE is an environment that uses hardware-enforced mechanisms to protect memory and execution from the operating system (OS) and its application layer (collectively known as the Rich Execution Environment, or REE \cite{trustzone}). 
Deploying  models in a TEE mitigates privacy-stealing attacks from REE-based adversaries: even if the REE is compromised, the adversary is limited to black-box access to the model, whereas models deployed in the REE are exposed to white-box attacks.

On the other hand, using TEEs on end devices face security and functionality challenges.  
While Intel Software Guard Extensions (SGX) has been deprecated on end-user devices \cite{sgxwiki}, Arm's TEE—commonly known as TrustZone—remains a widely adopted on-device solution, implemented in various mobile platforms (e.g., Qualcomm, Trustonic).  However, as Cerdeira \etal \cite{cerdeira2020sok} showed, TrustZone has been the target of high-impact attacks due to  its security vulnerabilities. The high privilege level of TrustZone, has led vendors to impose functional restrictions in an effort to reduce the attack surface, restrictions such as lack of support for GPU accessing \cite{sun2022leap}, and small memory size (32MB for OP-TEE) \cite{opteesize}.

To overcome these limitations, several solutions have proposed partitioning models and executing only the more sensitive components within TEEs \cite{mo2020darknetz, mo2021ppfl}. These approaches aim to provide near black-box security without placing the entire model inside the TEE. However, Zhang \etal \cite{zhang2024tsdp} demonstrate that such solutions remain vulnerable to privacy-stealing attacks and are not as secure as commonly assumed. Even partial model weights can leak private information about the training dataset, particularly when combined with publicly available resources (e.g., similar datasets or pre-trained models). Therefore, deploying the entire model within the TEE boundary remains the most effective strategy for protecting it against privacy attacks.

Arm Confidential Compute Architecture (CCA) \cite{ccasite} is a key component of the Armv9-A architecture that is expected to be available on Arm devices. Arm CCA allows the creation of special virtual machines called \textit{realm}, orthogonal to the already existing TrustZone.  
Realm is de-privileged as it has virtualized access to the resources, and it is TEE because it has protection against REE actors. 
Realm creation and runtime are supported by hardware-backed attestation services which can provide enough evidence for a relying party (\eg model provider) about the trustworthiness of the realm. Compared to TrustZone, CCA benefits from a more flexible memory allocation scheme.\footnote{TrustZone enforces isolation using an Address Space Controller (TZASC) and bus-level protection, requiring coarse-grained changes to memory regions. In contrast, CCA uses standard page tables and the Memory Management Unit (MMU) to enforce isolation, allowing for fine-grained and dynamic memory management.} The CCA features seem promising for on-device model deployment. Given that Arm is the dominant architecture in mobile and edge devices, we anticipate CCA's widespread deployment in the near future. 

\para{Motivation.} Inspired by (1) the limitations of existing TEE solutions, (2) vulnerabilities in current model partitioning strategies, and (3) the key features of Arm CCA, this work introduces and evaluates a framework for deploying on-device models within Arm CCA. We use the latest tools and plugins provided by Arm to simulate and trace Arm CCA behavior in running ML workloads. We do not employ partitioning strategies, ensuring that the entire model remains protected from REE actors. Our findings and evaluation results can be useful to support further research and early adoption, prior to the widespread adoption of CCA on end devices.
%
%

Our contributions are as follows:

\begin{itemize}
    \item We define a basic framework for on-device model deployment within Arm CCA and use the latest tools, software, and firmware to simulate the framework.
    \item We evaluate the framework for models of different sizes and applications, all showing acceptable overhead (22\% in the worst case).
    \item To showcase the security gain of the framework, we implement a membership inference attack on the models, showing that running models within a realm, on average, provides an 8.3\% decrease in the success rate of membership inference attacks against the training dataset.
    \item We make all our code and framework openly available and will maintain it to benefit early-stage adoption of CCA software products\footnote{\url{https://github.com/comet-cc/CCA-Evaluation}}.
\end{itemize}

%% file: parts/background.tex
In this section, we first provide a brief overview of Arm CCA (Section \ref{sec:armcca}) and why it comes with overhead (Section \ref{sec:realmoverhead}). We also discuss the possible choices of evaluating CCA (Section \ref{sec:CCAEVAL}). Finally, in Section \ref{sec:membershipbackground}, we introduce a privacy-stealing attack commonly used in security evaluations of ML systems.
\begin{figure}[tbp]
    \centering
    \includegraphics[width=1\linewidth]{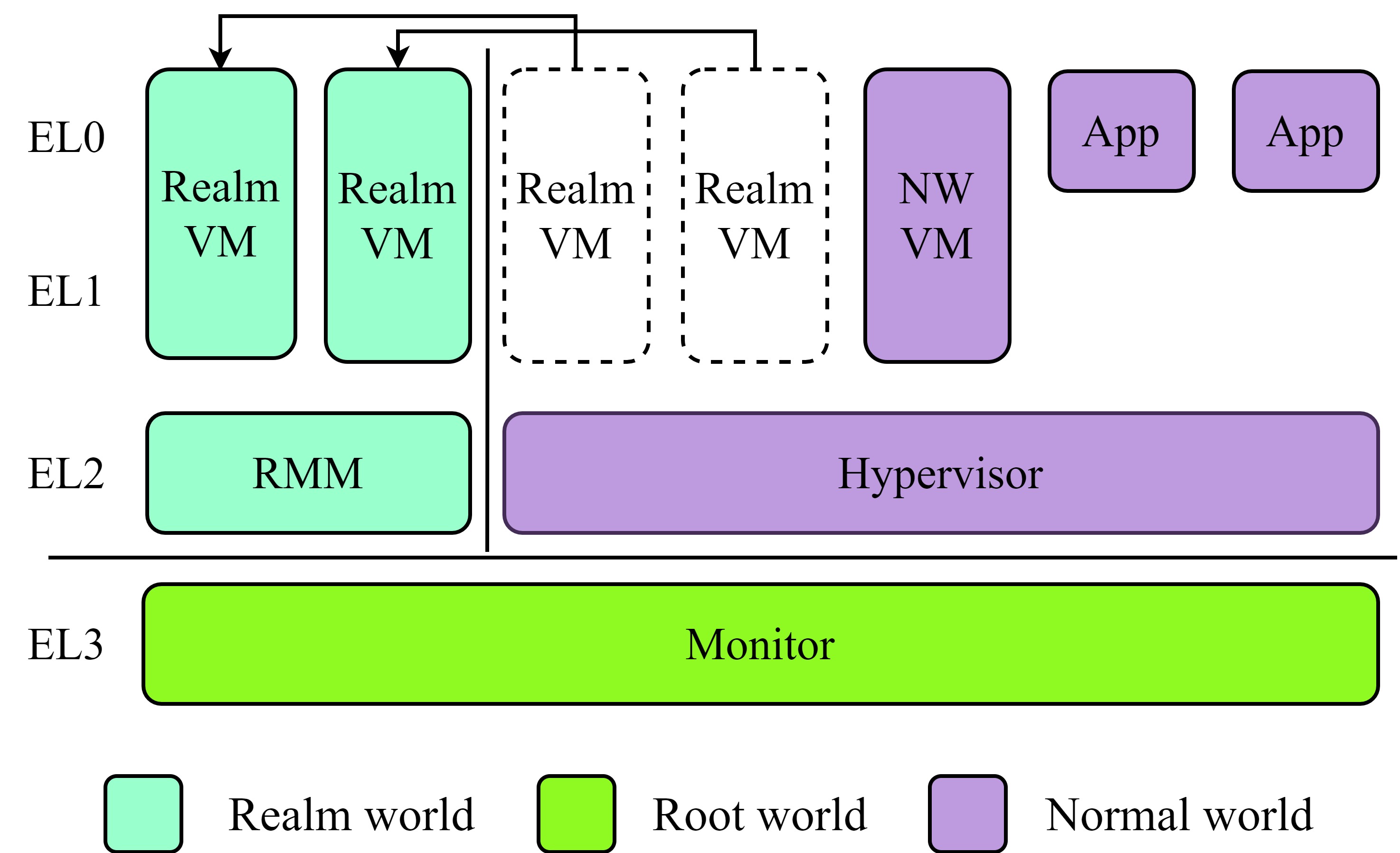}
   \caption{Arm CCA software architecture. The hypervisor allocates resources to realms but cannot access those resources, due to isolation boundaries between the realm and the normal world}
    \label{fig:Arm CCA software architecture}
\end{figure}
\subsection{Arm Confidential Compute Architecture}\label{sec:armcca}
Arm CCA \cite{ccasite} is a series of hardware and software architecture extensions that enhances Armv9-A  support for confidential computing. 
\begin{table}[tbp]
  \centering
\caption{Memory access rules applied by granule protection check (GPC)}
      \resizebox*{\linewidth}{!}{%
    \begin{tabular}{|l| c c c c|}
          \cline{1-5}
        \textbf{Security State} & \textbf{Normal PAS} & \textbf{Secure PAS} & \textbf{Realm PAS} & \textbf{Root PAS} \\
        \cline{1-5}
       Normal  & Yes & No  & No  & No\\
          Secure & Yes & Yes & No & No\\
         Realm & Yes & No &  Yes & No \\
         Root & Yes & Yes & Yes & Yes \\
       \cline{1-5}
    \end{tabular}
    }
    \label{tab:memoryaccess}
\end{table}
As shown in Figure \ref{fig:Arm CCA software architecture}, Arm CCA, introduces four worlds\footnote{In some references, \textit{execution environment} is used instead of \textit{world}, however in this work they are interchangeable.}: root, realm, secure, and normal world (a.k.a., non-secure world).  To enforce isolation between the worlds, CCA introduces a mechanism called granule protection check (GPC). Any memory access request succeeds only if the requester state (e.g., processor state) and the memory's state both comply with the rules defined in Table~\ref{tab:memoryaccess}.
Particularly, the root world state can access the physical address space (PAS) of all the other worlds while the realm and secure worlds state have access to the normal PAS, but they cannot access each other's PAS. 
The normal world (NW) state cannot access the PAS of the other worlds.
%
Arm architecture allows different exception levels (EL) to exist, from EL3 (the highest privilege) to EL0 (the lowest privilege).

\para{Software architecture.} Figure \ref{fig:Arm CCA software architecture} shows the software architecture of Arm CCA.  The Monitor is the highest privileged firmware in the system responsible for initially booting all EL2 firmware/software, managing the GPC, and context switching between different worlds. 
The normal world stack consists of a hypervisor operating at NW-EL2, virtual machines running at EL1 and EL0 and user-space apps running at EL0. The hypervisor is responsible for managing all resources (\eg CPU and memory) in the system.  
The realm world stack consists of a lightweight firmware known as Realm Management Monitor (RMM) which mediates resource allocation of realm VMs, and realm VMs (or simply realms) running at EL1 and EL0. RMM enforces isolation boundaries between the hypervisor and the realm VMs, making realm resources (\eg memory pages of the realm VM) inaccessible for the hypervisor. The RMM is also able to generate an \textit{attestation report} for the realm VM. This report keeps necessary information about the initial content of the realm as well as the firmware (RMM and Monitor) in the system \cite{sardar2023sok,RMM}.
\subsection{Realm Overhead}\label{sec:realmoverhead}
Handling exceptions\footnote{In Arm architecture, exceptions are conditions or system events that require action by privileged software \cite{armexception}. Notably, interrupts triggered by virtual devices (\eg virtual timer) are a type of exception.} is more complex for a realm VM, compared to a normal world VM. For a normal world VM, every exception is directly received and handled by the hypervisor, while for a realm VM, the RMM is initially responsible for handling the exception and, if necessary, forwarding it to the hypervisor. This complexity increases the overhead of running a workload within realm.
Two notable sources of exceptions are the hypervisor's timer interrupt (timer at NW-EL2) and the realm's timer interrupts (timer at Realm-EL1), both necessary for process scheduling within the two kernels. Each time these timers are acknowledged by the processor, an exit from the realm occurs, which requires handling by the hypervisor.  In Section \ref{sec:inferenceoverhead} and Appendix \ref{section:Inference Overhead appendix} we compare the runtime execution and I/O operation between a realm and a NW VM.
\subsection{CCA Evaluation Platforms} \label{sec:CCAEVAL}
At the time of writing, there is no hardware compatible with the CCA specification.  However, there are software that emulates the behavior of a CCA-compatible device. Linaro's QEMU \cite{QEMUlinaro} can be used to boot and run the CCA software stack \cite{QEMURME}. Fixed Virtual Platform (FVP) is the official software released by Arm, compatible with the CCA specification \cite{armccaenablement,fvp}. FVP provides useful plugins and tools which, combined, provide detailed information about the behavior of CCA. Devlore \cite{bertschi2024devlore} used QEMU, but other works have used FVP in their evaluation  as functional prototype \cite{wang2024cage,zhang2023shelter} and also performance prototype \cite{sridhara2024acai,chen2024cubevisor,siby2024guarantee}. We utilize FVP\footnote{More specifically, we use FVP\_Base\_RevC\-2xAEMvA\_11.25\_15} for the evaluation. In Appendix \ref{section:Experimental Setup}, we describe how FVP can be set up with plugins and tracing tools to measure realm's behavior. Furthermore, we explain the accuracy of FVP and other possible options to evaluate CCA. It is important to note that neither FVP nor QEMU is designed to provide performance predictions, and any evaluation based on these tools should be regarded as preliminary and approximate.
%
\subsection{Membership Inference Attack}\label{sec:membershipbackground}
Membership inference attacks (MIA) are a class of attacks in which an adversary tries to determine whether a particular data point was a part of the training set or not.  These attacks have been widely used in the literature to assess how much a system ``leaks'' information about the training dataset \cite{mo2020darknetz,mo2021ppfl,zhang2024tsdp}.  
%
In a typical attack setting \cite{liu2022ml,zhang2024tsdp}, an adversary has access to a \textit{shadow dataset} which is statistically similar to the target model's dataset. This dataset is then used to train a \textit{shadow model} and an \textit{attack model} (a binary classifier). Finally, to determine whether a data sample is a member of the target model’s training dataset, the sample is fed to the target model, and the posteriors and the
predicted label (transformed to a binary indicator on whether
the prediction is correct) are fed to the attack model. Moreover, an adversary with white-box access can enhance the attack's accuracy by leveraging additional model information, such as classification loss and sample gradients (see \cite{liu2022ml,zhang2024tsdp} for details). We use this attack in Section \ref{sec:membershipinference} to show the privacy protection of our framework.

%% file: parts/model.tex
In this section, we describe a basic framework to deploy on-device models within CCA. We follow the system model introduced by \cite{siby2024guarantee}.

\subsection{System Model} \label{sec:systemmodel}
As illustrated in Figure \ref{fig:ArmCCAsoftwarearchitecture}, the system involves three parties: \textit{model providers}, \textit{clients}, and a \textit{trusted verifier}.
A model provider is an entity responsible for training and deploying ML models on end-devices for various tasks. These models, along with their training datasets, are considered intellectual property and must be protected from unauthorized access by malicious users and other model providers.
The client is an end-device, such as a smartphone or an IoT gateway, which supports Arm CCA. Clients host a wide variety of applications within their REE that may require machine learning services, such as facial recognition, voice detection, or chat assistants.
The trusted verifier is responsible for providing realm images. A realm image includes a complete stack for a virtual machine, encompassing an operating system, user-space libraries, and applications necessary for running the model within the realm.
\subsection{Threat Model} \label{sec:threatmodel} 
We assume that model providers and clients are two mutually distrusting entities, but they both trust the images offered by the trusted verifier. Clients may attempt to maliciously extract information about the model's weights and training data.
 Both the Monitor and the RMM are considered trustworthy due to their small codebase, and formal verification in the case of the RMM ~\cite{li2022design,fox2023verification}. However, the NW stack is untrusted as it is large and complex, containing unverified user-space applications, third-party libraries, and drivers. An adversary could exploit these vulnerabilities to compromise the entire NW. 
Arm CCA, by default, does not provide availability guarantees regarding runtime execution and memory of realm. However, we assume that the hypervisor allocates sufficient CPU time and memory to the realm, allowing it to effectively load the model and perform inference\footnote{Altering these assumptions does not impact the security of the model, it only affects the quality of the ML service experienced by the NW app.}.
Physical and side-channel attacks are also significant threats to the deployment of the device model \cite{yuan2024ciphersteal,yuan2024hypertheft}. However,  there are considered out of scope and the hardware is trusted.
%

%% file: parts/system.tex
\subsection{Model Deployment Pipeline}
\begin{figure}[t!]
    \centering
    \includegraphics[width=1\linewidth]{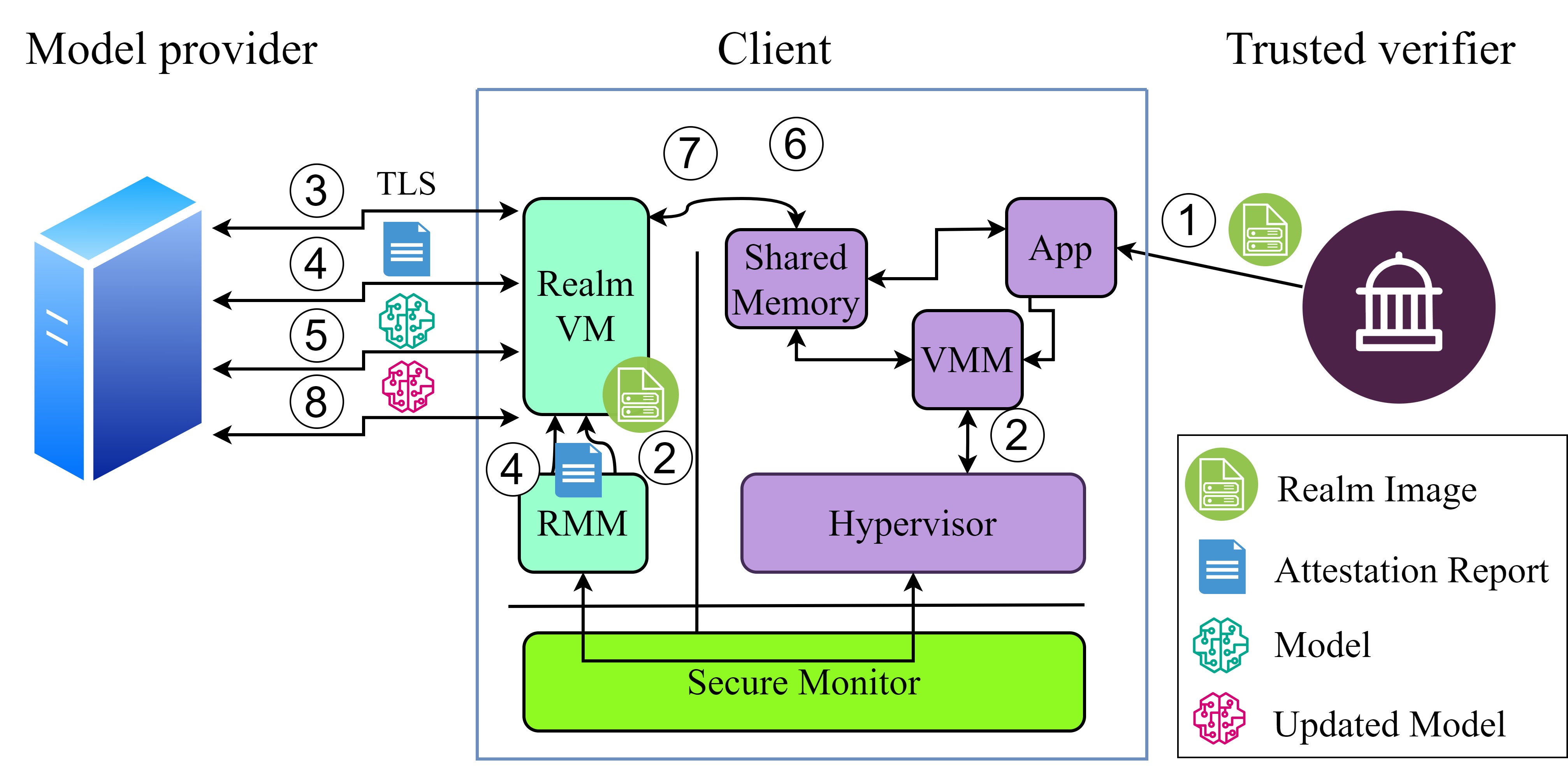}
      \caption{Overview of the steps required for running a ML model on the client edge device. We show a simplified view of the normal and realm worlds within the client. The client's steps are (1) obtaining realm image from verifier (2) creating and activating a realm VM (3) establishing connection with provider (4) realm attestation (5) obtaining model from provider (6) announcing model readiness to normal world (7) running inference (8) performing model updates.}
    \label{fig:ArmCCAsoftwarearchitecture}
\end{figure}

Figure~\ref{fig:ArmCCAsoftwarearchitecture} shows an overview of our framework. In the following, we provide a description of the various steps involved in deploying the model within a realm.

\para{Realm setup.} A NW app starts the process by obtaining a publicly-available and verified realm image from the trusted verifier (Step 1 in Figure~\ref{fig:ArmCCAsoftwarearchitecture}). The realm creation is done by a collaboration among a virtual machine manager (VMM) at NW-EL0, the hypervisor, and the RMM (Step 2). After populating the realm memory, the hypervisor sends the activation command to the RMM. 
Once the realm is activated, it can receive CPU time, and the hypervisor is no longer able populate new content into the realm's address space.

\para{Model initialization.} 
After realm's kernel is booted, the realm establishes a TLS connection with the model provider (Step 3). 
Later, the realm sends an attestation request to the RMM and in return, the RMM sends the attestation report to the realm, which is forwarded to the model provider (step 4).
The model provider can now use  the attestation report to verify the content of the realm and decide whether it can trust the realm or not. 
On verification, the model provider sends the model to the realm via the TLS connection (Step 5). 

\para{Inference.} 
The realm's kernel includes a virtio-9p driver, which is used to establish a file-system-based shared memory with the NW app. 
After receipt of the model, the realm announces its ability to respond to inference queries to the NW app (Step 6).  
Later, the NW app sends input data to the realm, the realm feeds it into the model, obtains the inference, and writes the output back to the shared file system so that the NW app can read it (Step 7).

\para{Service maintenance.} 
In addition to performing inference, the framework must also handle other maintenance operations. For example, a model provider might set usage limits—such as a validity period or maximum inferences—by embedding this functionality in the realm image. Once these limits are reached, the realm calls the hypervisor to terminate and release its memory. The realm can also periodically query the model provider for updates on the model (step 8).

\para{Integration with mobile devices.} Our framework can be adapted for deployment on mobile devices. A potential setup involves a hypervisor supporting CCA running at EL2, with Android at EL1 and user applications at EL0. In this configuration, while Android remains responsible for managing applications running at EL0, the hypervisor can create and manage Realm VMs, enabling secure execution environments for sensitive models.

%% file: parts/Evaluation.tex
In this section, we evaluate and compare our framework against a baseline scenario in which the model is deployed within a NW VM. We show the computational overhead and privacy benefit of our framework.
\subsection{Experimental Setup}\label{Experimental Setup}
In this section, we describe the experimental setup  used to  evaluate our framework. We compare our framework with a baseline scenario, involving deploying model within a normal world VM. We use FVP to report the overhead of our framework in comparison to the baseline. FVP is instruction-accurate, that is, it accurately models the instruction-level behavior of a real processor that supports CCA \cite{FastModel,fvp}. However, it does not effectively capture certain micro-architectural behaviors (e.g., caching and memory accesses), which makes cycle-accurate and timing-based measurements unreliable \cite{FastModel}. While we use FVP to report the number of instructions executed by the FVP's processor core, these measurements should be regarded as preliminary estimations. We do not claim that they represent the actual performance overhead on real CCA hardware. In Appendix \ref{section:Experimental Setup}, we provide extensive information on how to set FVP in conjunction with tracing tools to accurately measure number of instructions executed by the FVP's core. We also use \textit{Shrinkwrap} \cite{shrinkwrap}, a tool that simplifies the building and execution of firmware/software on FVP. Shrinkwrap automatically downloads and builds necessary firmware based on the given configuration files. More information on software and firmware version we used for the evaluation is provided in Appendix \ref{section:Experimental Setup}.

\para{Privacy protection.}
As discussed in the threat model (Section \ref{sec:threatmodel}), all software in the normal world is considered untrusted. Consequently, in the baseline scenario—where the model runs within a normal world VM—the model is entirely exposed to potential adversaries. An adversary with this level of white-box access can launch privacy-stealing attacks to infer information about the model’s training dataset.
In contrast, our framework protects the model by executing it within realm, effectively concealing its weights from NW adversaries. Specifically, our framework restricts the adversary’s access to the model to a black-box setting, where only query access is allowed. In Section \ref{sec:membershipinference}, we demonstrate the resulting privacy advantages by evaluating both white-box and black-box membership inference attacks.

\begin{table*}[htbp]
  \centering
\caption{Experimental settings used in the evaluation. The VM size  depends on runtime memory use of inference code, size of model, and size of  dynamic libraries required for each model.}
 \begin{adjustbox}{max width=\textwidth}
  \begin{tabular}[c]{|cccccc|}
  \cline{1-6}
  \textbf{Experimental Setting} & \textbf{Model} & \textbf{Model Size (MB)} & \textbf{Library (API)} & \textbf{Input Format}  & \textbf{VM size (MB)} \\
 \cline{1-6}
    \circled{1} & AlexNet  & 9  & TensorFlow Lite (C++)  & .bmp & 300 \\
     \circled{2} & MobileNet\_v1\_1.0\_224 \cite{litemobilenet}  & 16  & TensorFlow Lite (C++)  & .bmp & 400 \\
     \circled{3} & ResNet18  & 44  & TensorFlow Lite (C++)  & .bmp & 450 \\
     \circled{4} & Inception\_v3\_2016\_08\_28 \cite{tensorflowinception}  & 95 & TensorFlow (C++) & .jpg  & 1750 \\
       \circled{5} & VGG  & 261 & TensorFlow (C++) & .wav  & 3650 \\
       \circled{6} &  GPT2 \cite{gpt2}  & 177 & llama.cpp \cite{llamacpp} (C++)  &   text  & 900 \\
         \circled{7} &   GPT2-large \cite{gpt2-large}  & 898 & llama.cpp \cite{llamacpp} (C++)  & text    &  1800 \\
 \circled{8} & TinyLlama-1.1B-Chat-v0.5  \cite{TinyLlama} & 1169 & llama.cpp \cite{llamacpp} (C++)  & text & 2000 \\
\cline{1-6}
  \end{tabular}
  \end{adjustbox}
  \label{tab:models}
\end{table*} 
\para{Models.} Table~\ref{tab:models} shows an overview of the models and settings used in the evaluation. We choose models of varied sizes and types for typical on-device tasks like image classification, speech recognition, and chat assistants. For each model, an appropriate VM size is chosen, which is enough for the run-time progress of inference. 
The size of the virtual machine depends mainly on the use of inference code, the size of the model, and the size of dynamic libraries required for each model.  
\subsection{Inference Overhead}\label{sec:inferenceoverhead}
\begin{table*}[t]
    \centering
    \caption{Mean (standard deviation) of instructions executed per inference service. Each experimental setting is described in Table \ref{tab:models}. }
    \begin{adjustbox}{max width=\textwidth}
    \begin{tabular}{|c|ccc|ccc|ccc|ccc|ccc|}
        \cline{1-16}
        \multirow{2}{*}{\textbf{Setting}}  & \multicolumn{3}{c|}{\textbf{Model Initialization ($10^6$)}}  & \multicolumn{3}{c|}{\textbf{Read Input ($10^6$)}} & \multicolumn{3}{c|}{\textbf{Inference Computation ($10^6$)}}  & \multicolumn{3}{c|} {\textbf{Write Output ($10^6$)}} &  \multicolumn{3}{c|}{\textbf{Total ($10^6$)}}  \\
        \cline{2-16}
       & \textbf{\textit{R VM}} & \textbf{\textit{NW VM}} & \textbf{\textit{Ovh}} & \textbf{\textit{R VM}} & \textbf{\textit{NW VM}} & \textbf{\textit{Ovh}} & \textbf{\textit{R VM}} & \textbf{\textit{NW VM}} & \textbf{\textit{Ovh}} & \textbf{\textit{R VM}} & \textbf{\textit{NW VM}} & \textbf{\textit{Ovh}}  & \textbf{\textit{R VM}} & \textbf{\textit{NW VM}} & \textbf{\textit{Ovh}} \\
        \cline{1-16}
        \circled{1} & 1.6  & 1.2  & 33\% & 0.6   & 0.3   & 100\% & 98.0   & 82.0  & 19\% & 1.1  & 0.5   & 120\% & 105.9  & 87.8  & 20\%     \\
       \circled{2}  & 1.7  & 1.2  & 41\% & 4.7   & 1.1 & 100\% & 335.4  & 278.9   & 20\% & 0.7  & 0.3   &  133\% & 351.8  & 289.3  & 21\%\\
       \circled{3}  & 2.1  & 1.6   & 31\%   & 0.6   & 0.3  &  100\% & 418.2   & 344.0  &  21\%  &  0.9  & 0.4   & 125\%  & 442.8  & 363.2   & 20\%\\
       \circled{4}    & 397.9  & 333.4  & 19\% & 2.8    &  1.8  & 55\% & 7663.8     & 6382.8    & 20\% & 4.6   & 3.5  &  31\% & 8717.2  & 7201.1  & 21\%  \\
      \circled{5}  & 345.1  & 295.8  & 16\% & 1.8  & 1.1  & 63\% & 6365.7  & 5420.7  &  17\% &  0.15  & 0.09  & 66\% & 6713.2 & 5717.9   & 17\% \\
      \circled{6}  & 1039.1  & 821.9  & 26\% & 2.7  & 1.8  & 50\% & 12036.6   & 9858.7     & 22\% & 0.11  & 0.04   & 75\% & 13144.9  & 10726.3   & 22\% \\
        \circled{7}  &  2653.6   & 2158.5  & 22\%  & 2.7  & 1.8   & 50\% & 73603.1   & 59870.6    & 22\% & 0.07  & 0.04    & 75\%   & 76412.3  & 62156.4   & 22\% \\
           \circled{8} & 2784.9  &  2312.1   & 20\%  &  2.6  & 1.8  & 44\% & 94480.0   & 79452.7     & 18\% & 0.07   &  0.04    & 75\% & 97433.3  & 81905.6  & 18\%\\
        \cline{1-16}
    \end{tabular}
    \end{adjustbox}
    \label{tab:evaluationinference}
\end{table*}
In order to evaluate the overhead of our framework, we perform an evaluation with two scenarios. In the baseline scenario, the model and the code are stored in a NW VM.
In the second scenario, the model and code are stored in a realm VM. In both scenarios, a file system-based data sharing is established between the VM and the NW app, allowing the NW app to send input queries and receive inference outputs. In order to get more insights about the inference service, we divide the service into four stages and measure each one separately, (1) model initialization, which involves loading the model into memory allocated by the inference code, (2) getting input from the NW app and storing it in the inference code memory (3) inference computation, which refers to local computations within the VM to obtain the output, and (4) writing the output back to the NW. 
For each experiment in both scenarios, we instantiate five VMs and perform five inferences per VM, yielding a total of 25 repetitions per configuration. We report the mean values, however standard deviations are omitted as they are consistently below 10\% in all experiments.

Table \ref{tab:evaluationinference} shows the results of our evaluations. As illustrated, the total overhead of inference service within the realm is moderate, ranging from 17\% to 22\%.
Model initialization overhead varies between 16\% to 41\% depending on the API used for inference. The highest numbers are within  experiments  \circled{1}, \circled{2}, and \circled{3}, all using the TFlite API. On the other hand, overhead of read input and write output are between 44\% and 100\%, and 31\% and 133\%,  respectively, showing considerably bigger overhead in I/O-involved operations within realm. The variation in input read overhead is primarily due to differences in input size across models, while the variation in output write overhead is attributed to the number of output classes and the format in which outputs are returned to the NW app.
As explained in Section~\ref{sec:realmoverhead}, the main contributor to these overheads is the increased complexity of exception handling in the realm. Although I/O operations are relatively expensive in this setting, they represent only a small portion of the total computation and do not significantly affect the overall inference performance. We also perform another experiment to see how much each entity is responsible for the overhead of inference computation and report the results in Appendix \ref{section:Inference Overhead appendix}. Finally, it is important to note that these results represent only an initial approximation, based on the number of instructions executed by the simulator’s core. We do not report these figures as overheads that would be replicated on actual CCA hardware.
\begin{table*}[t]
    \centering
    \caption{Adversary's success rate in the membership inference attack. 
 NW: Model is deployed within NW, giving the adversary white-box access to the model,
 RW: Model is deployed within realm world, giving the adversary black-box (label-only) access to the model. R is the ratio between the size of shadow dataset and the size of training dataset}
    \label{tab:membershipattack}
\begin{tabular}{|c|c|ccc|ccc|ccc|c|}
\hline
\multirow{2}{*}{\textbf{Model}}  &  \multirow{2}{*}{\textbf{Deployment}} &  \multicolumn{3}{c|}{\textbf{R = 1}} & \multicolumn{3}{c|}{\textbf{R = 1/4}} & \multicolumn{3}{c|}{\textbf{R = 1/8}} & \multirow{2}{*}{\textbf{Total (Average)}} \\ 
\cline{3-11} 
 &  & \textbf{\textit{CF10}} & \textbf{\textit{CF100}} & \textbf{\textit{CelebA}} & \textbf{\textit{CF10}} & \textbf{\textit{CF100}} & \textbf{\textit{CelebA}} & \textbf{\textit{CF10}} & \textbf{\textit{CF100}} & \textbf{\textit{CelebA}} & \\ 
 \cline{1-12} 
\multirow{3}{*}{AlexNet} & NW & 71.8 & 84 & 84.9 & 65.4 & 83.9 & 82.4 & 57.3 & 76.9 & 82.6 & \\ 
 & RW & 68.9 & 76.0 & 69.0 & 66.3 & 50.0 & 68.6 & 67.9 & 50.0 & 60.6 & \\ 
 \cline{2-12} 
 & Diff & 2.9 & 8.0 & 15.9 & -0.9 & 33.9 & 13.8 & -10.6 & 26.9 & 22.0  & 111.9 (12.4) \\ 
 \cline{1-12} 
\multirow{3}{*}{ResNet18} & NW & 70.0 & 91.9 & 84.1 & 69.9 & 89.4 & 86.9 & 66.4 & 87.9 & 85.5 & \\  
 & RW & 68.9 & 85.8 & 81.4 & 68.5 & 73.3 & 81.0 & 68.9 & 80.9 & 80.0 &  \\ 
 \cline{2-12} 
 & Diff & 1.1 & 6.1 & 2.7 & 1.4 & 16.1 & 5.9 & -2.5 & 7.0 & 5.5  &  37.8 (4.2)\\ 
  \cline{1-12} 
\end{tabular}
 
\end{table*}

\subsection{Membership Inference Attack}\label{sec:membershipinference}
To demonstrate the security benefits of our framework, we conduct both white-box and black-box membership inference attacks on two models (experimental setting \circled{1} and \circled{3} in Table \ref{tab:models}). 
We adopt the MIA proposed in \cite{zhang2024tsdp}, using the same settings and  hyper-parameters (\eg learning rate, number of epochs, etc). In this attack, the adversary has access to a shadow dataset drawn from the same distribution as the training dataset. The adversary then uses the shadow dataset to train a binary classifier that infers membership in the target training dataset (see Section \ref{sec:membershipbackground} for details).
While the default assumption in \cite{zhang2024tsdp} is that the shadow dataset size matches that of the training dataset, this assumption may not be realistic in all practical scenarios. To account for this, we experiment with three different ratios between the training dataset and shadow dataset sizes. Specifically, we fix the size of the training dataset across all scenarios and reduce the shadow dataset size to 1/4 and 1/8 of the training dataset size. We conduct these experiments using two models and three different datasets.
The results, presented in Table \ref{tab:membershipattack}, shows that the adversary’s success rate decreases by an average of 12.4\% and 4.2\% for AlexNet and ResNet18, respectively (8.3\% reduction on average). These findings are consistent with similar observations in \cite{liu2022ml}. Notably, the gap between the adversary's success rates in the two settings grows as the number of output classes increases. The gap is larger for CIFAR100 (100 output classes) than for CIFAR10 (10 output classes) and CelebA (configured for 32 output classes in our evaluation).

%% file: parts/discussion.tex
\para{Realm device assignment.}
Device assignment is one of the planned future enhancements for CCA \cite{armCCAupdate}, and it could enable the deployment of new capabilities across the ML pipeline. Securely assigning specialized hardware—such as GPUs and NPUs—to realm  could significantly accelerate inference computation. More importantly, device assignment opens the possibility of protecting the entire inference pipeline within the TEE boundary. Although our current system protects the model itself from NW adversaries, it does not protect the source of input data. In safety-critical applications—such as health monitoring or autonomous driving—corrupted inputs can pose serious risks. Thus, achieving strong guarantees requires securing the entire inference workflow, including:(1) input generation, (2) delivery of inputs to the model, (3) generation of outputs, and (4) consumption of outputs by the requester.

\para{Membership Attack on LLMs.}
The larger memory size of the realm, as compared to other on-device TEE solutions, allowed us to run LLMs within a realm. However, we did not show the privacy benefit of running the LLM within a realm. Future works could explore the trade-off between performance and privacy when deploying LLMs in realm compared to NW. Currently, several studies have examined MIA in black-box settings \cite{galli2024noisy, xie2024recall} while others~\cite{duan2024membership} have questioned the assumptions of previous attacks, investigating whether MIAs are feasible under more realistic conditions for LLMs. White-box MIAs for LLMs remain an emerging area, with no proposed white-box attacks demonstrating consistent superiority over black-box approaches.

\para{Limitation.} 
For the evaluations in this paper, we have emulated CCA using FVP, our results are only initial approximation not obtained from a real hardware.  
Accurate evaluation can be done in the future when a real device supporting Arm CCA will be available.

%% file: parts/relatedworks.tex
\para{Model partitioning on end-devices.}
 To overcome limitation of current TEEs, several works have proposed to partition model in which more sensitive parts are running within a TEE -- these include shielding deep layers \cite{mo2020darknetz,mo2021ppfl}, shallow layers \cite{hou2021model}, intermediate layers \cite{shen2022soter}, non-linear layers \cite{sun2023shadownet} within a TEE. 
Zhang \etal \cite{zhang2024tsdp} showed that those partitioning solutions are vulnerable to privacy attacks when public information like datasets and pre-trained models engages in attacks. 

\para{TEE extensions.}
There are works aim to overcome the limitations of TEE by introducing system-level techniques. SANCTUARY \cite{brasser2019sanctuary} and LEAP \cite{sun2022leap}, for instance, create isolated user-space enclaves in NW  on top of TrustZone. However, in both works, the secure world (TrustZone) is trusted, making them vulnerable to malicious actors within the secure world. This is a significant concern, as \cite{cerdeira2020sok} demonstrated, current implementations of TrustZone suffer from critical vulnerabilities. To address these issues, REZONE \cite{cerdeira2022rezone} proposes a system that de-privileges the TEE's operating system, offering enhanced protection against potentially malicious TEE components.  Li \etal \cite{li2024teem} Introduces a method to allocate large memory for  TrustZone apps by modifying OP-TEE.  However, the total amount of memory available to OP-TEE remains limited to the configured size at boot time.

\para{Systems based on CCA.} 
As CCA is still under development, there is limited prior work in this space. 
Formal methods is introduced in \cite{li2022design,fox2023verification} to verify security and functional correctness of RMM. 
 SHELTER \cite{zhang2023shelter} provides user-space isolation in the normal world using CCA hardware primitives. 
 ACAI \cite{sridhara2024acai} is a system that allows CCA realms to securely access PCIe-based accelerators with strong isolation guarantees. 
 DEVLORE \cite{bertschi2024devlore} is a system that allows realm VM to access legitimate integrated devices (\eg keyboard) with necessary memory protection and interrupt isolation from an untrusted hypervisor.
GuaranTEE \cite{siby2024guarantee} took initial steps in using CCA for ML tasks. This framework provides attestable and private machine learning on the edge using CCA and evaluated it for running a small model within realm.  In this work, we adopt their system model and utilize tracing tools to estimate the system's overhead.

%% file: parts/conclusion.tex
In this paper, we presented an in-depth evaluation of Arm's Confidential Computing Architecture (CCA) as a solution to protect on-device models.
We measure both the overhead and the privacy gains of running models of various sizes and functionalities within a realm VM. 
Our results indicate that, CCA can be a viable solution for model protection. 
While various challenges still remain before CCA's widespread deployment, we provide the first indication of its suitability as a mechanism to provide model protection.

%% file: parts/Acknowledge.tex
We wish to acknowledge the thorough and useful feedback from anonymous reviewers and our shepherd. The research in this paper was supported by the UKRI  Open Plus Fellowship (EP/W005271/1 Securing the Next Billion Consumer Devices on the Edge) and an Amazon Research Awared “Auditable Model Privacy using TEEs”.

%% file: parts/FVP_setup.tex
\para{Software stack.}  We use the Trusted Firmware-A~\cite{TF-A} (v2.11), and the Trusted Firmware implementation of RMM~\cite{RMM} (tf-rmm-v0.5.0) as the Monitor and the RMM of the software stack (Figure \ref{fig:Arm CCA software architecture}), respectively. We separately build linux-cca \cite{linux-cca} and the file system for each experiment and pass them to Shrinkwrap.
Shrinkwrap later boots  FVP with the necessary firmware and the given kernel and file system. 
We also use Buildroot \cite{buildroot}, to create customized file systems for each experimental setup.  
In order to create a virtual machine, we need to provision a virtual machine manager (VMM) to the hypervisor's file system. Both kvmtool-cca \cite{kvmtool-cca} (cca/rmm-v1.0-eac5) and Linaro's QEMU \cite{QEMUlinaro} (cca/v3) have support for realm VMs, but for each one, we need to use a compatible branch of linux-cca (which has a similar name to the branch of that VMM). 

 \para{FVP Accuracy.} FVP promises to accurately model the instruction behavior of a real processor \cite{FastModel,fvp}. However, some micro architectural behaviors (\eg caching and memory accesses) are different between FVP and an actual device, making cyclic and timing measurements unreliable. \cite{FastModel}. Therefore, we do not report timing or cycle-accurate performance results from the simulation.
While some studies \cite{zhang2023shelter,sridhara2024acai,wang2024cage} have prototyped CCA on existing Armv8-A hardware, these platforms lack essential features—such as GPC support in system registers and accurate cache behavior—which pose challenges to the accuracy of such prototypes. Although our framework is evaluated using FVP, these hardware-based prototypes may still be valuable for others, particularly for enabling cycle-level and timing evaluations.

\para{Instruction tracing in FVP.} FVP can be used in conjunction with tracing tools and plugins to provide detailed information about the behavior of CCA. Particularly, we use \textit{GenericTrace}  to choose a trace source (\eg instructions in our case) and \textit{ToggleMTIPlugin} to enable/disable tracing during runtime. We configure GenericTrace to trace and print each instruction executed by an FVP's processor core, along with other metadata. The metadata includes the security state and the exception level of the core when running that instruction, and the total number of instructions executed until that point in time. 
Using ToggleMTIPlugin, FVP can be set to be sensitive to a particular assembly instruction\footnote{We used HLT 0x1337}. Whenever this instruction is executed by the FVP's processor core, tracing is automatically started/stopped. We add this instruction at points in the code to enable and disable GenericTrace. This is necessary to  reduce the size of the trace file and only get what it is necessary for each experiment.
 Lastly, similar to what has already been done by Sridhara \etal~\cite{sridhara2024acai}, we add a set of assembly instructions to the code to mark specific points (\eg beginning and end of inference) in the final trace file. Later, by analyzing this trace file, we can get the result of evaluations including number of instructions executed (for example between the beginning and end of inference).

\para{Runtime isolation.} Since FVP simulates a multi-core device, additional measures are necessary to ensure that the target workload is executed exclusively on the traced core. To achieve this, we utilize a kernel-command line parameter called \textit{isolcpus} to isolate one core from the hypervisor’s general load balancing and scheduling algorithms. This ensures that the hypervisor's scheduler does not assign any processes to the traced core by default. Subsequently, during runtime, we use the \textit{taskset} tool to explicitly direct the hypervisor to use only the isolated core for the process that oversees the virtual machine.

\para{On-demand memory delegation.}
 During the VM's boot process, the hypervisor \cite{linux-cca} delegates only the physical pages necessary to load the kernel and file system images. The remaining memory in the VM's address space is delegated on-demand, triggered by the first access to those addresses. To decouple this one-time overhead from the main experiment in each evaluation, we address it by running a user-space program within the VM. This program temporarily allocates all available memory in the virtual machine’s user space and fills it with binary 1's. This ensures that the hypervisor delegates the entire memory beforehand, preventing any memory delegation during the main experiment.

\para{Experimental Hosts.} The membership inference attack in Section \ref{sec:membershipinference} is conducted on a system with dual Intel Xeon Gold 6136 CPUs (48 cores, 3.7 GHz max) and 251 GiB RAM, utilizing an NVIDIA Quadro GV100 GPU for acceleration. The environment run on Ubuntu 22.04.1 with kernel 6.5.0. Although FVP results are independent of the host platform, we report the system specifications for completeness. We conduct all FVP-related experiments on a Lenovo ThinkCentre M75t Gen 2 with 16GB RAM and an 8-core AMD Ryzen 7 PRO 3700 processor (OS: Ubuntu 22.04.4 LTS). We set FVP to have two clusters, each with four cores supporting Armv9.2-A and 4GB of RAM. 

%% file: parts/Inference_Overhead.tex
\begin{table*}[t]
    \centering
    \caption{Number of instructions (in millions) executed by each software/firmware component for a single inference in both normal and realm VMs. These results correspond to experimental setting 2 in Table \ref{tab:models}.}
    \begin{tabular}{|c|c|c|c|c|}
        \cline{1-5}
         \textbf{Exception} & \multicolumn{2}{c|}{\textbf{Realm VM Experiment}} &  \multicolumn{2}{c|}{\textbf{NW VM Experiment}}  \\
         \cline{2-5}
         \centering \textbf{Level} & \textbf{\textit{Realm World}} & \textbf{\textit{Normal World}} & \textbf{\textit{Realm World}} & \textbf{\textit{Normal World}} \\
         \cline{1-5}
       EL0  & 240.14 & 0.04 & 0 & 240.18  \\ 
       EL1  & 24.68 & 0 & 0 & 23.85  \\
              EL2 & 41.18 & 16.84 & 0 & 14.80  \\
        \cline{1-5}
              EL3  & \multicolumn{2}{c|}{5.13}   & \multicolumn{2}{c|}{0}  \\
        \cline{1-5}
    \end{tabular}
    \label{tab:perlevelinst}
\end{table*}

In order to identify the source of overhead within the inference computation, we conduct an additional experiment to quantify the engagement of firmware and software components during inference computation. Using configuration \circled{2}, we deploy two VMs -- one within the Realm world and the other in the NW -- with both performing the same task (a single inference). We then measured the number of instructions executed by each software and firmware component in the system. The results are presented in Table \ref{tab:perlevelinst}.
In both experiments, the number of executed instructions at EL0 and EL1 are relatively the same. However, significant differences emerge at EL2 and EL3, which are the main contributors to the overhead in the realm. Specifically, the virtualization support for the NW VM requires only 14.8 million instructions executed by the hypervisor. In contrast, the Realm VM required 16.84 million instructions executed by the hypervisor, with an additional 41.18 million instructions executed by the RMM and 5.13 million by the Monitor. These results suggest that the RMM is the main source of overhead, accounting for more than twice the number of executed instructions by the hypervisor. Worth noting that these measurements are done during the inference computation and there is no I/O involved.

%% file: parts/realm_setup_overhead.tex
\begin{table*}[t]
    \centering
    \caption{Mean (standard deviation) of number of instructions executed for realm boot and termination. Each experimental setting is described in Table \ref{tab:models}. }
    \begin{adjustbox}{max width=\textwidth}
    \begin{tabular}{|c|ccc|ccc|}
        \cline{1-7}
        \multirow{2}{*}{\textbf{Experimental Setting}}  & \multicolumn{3}{c|}{\textbf{VM Boot ($10^6$)}}  & \multicolumn{3}{c|} {\textbf{VM Termination ($10^6$)}} \\
        \cline{2-7}
          & \textbf{\textit{Realm VM}} & \textbf{\textit{NW VM}} & \textbf{\textit{Overhead}}  & \textbf{\textit{Realm VM}} & \textbf{\textit{NW VM}} & \textbf{\textit{Overhead}}  \\
        \cline{1-7}
        \circled{2}  & 7630.1 (52.6)  & 788.7 (0.7) & 867\% &   619.9 (3.3) & 83.3 (0.1)  & 644\%  \\
      \circled{4}   & 24960.7 (132.9)   &  1246.6 (0.9) & 1902\% &    2332.4 (2.4)   & 93.1 (0.2) &  2405\% \\
       \circled{5}  & 44499.3 (10.9) & 2329.4 (5.2) &  1832\%  &  5156.4 (6.9) & 142.4 (0.3) & 3521\% \\
       \circled{6} & 21101.5 (71.4) & 1195.0 (0.2) & 1665\%   & 1325.3 (2.4) & 87.1 (0.1)  & 1421\% \\  
        \cline{1-7}
    \end{tabular}
    \end{adjustbox}
  \label{tab:evaluationsetup}
\end{table*}

\label{sec:realmsetupverhead}
In this section, we evaluate the overhead associated with booting and terminating a realm VM in comparison to a baseline scenario (a NW VM). As illustrated in Table \ref{tab:evaluationsetup}, the overhead for booting and terminating a realm VM is substantial, with observed increases ranging from 867\% to 21,902\% for booting and from 644\% to 3,521\% for termination. These elevated overheads are primarily due to the additional RMM checks and processes required for page delegation (during boot) and reclaming those pages (during termination).
Notably, the overhead for both booting and termination escalates with the size of the VM, as reflected in the experimental settings detailed in Table \ref{tab:models}, with the exception of boot overhead between \circled{4} and \circled{5}. This results suggests that, although realm booting and termination represent one-time costs, they become significantly burdensome when deploying larger models, which typically necessitate larger VM sizes.